\documentclass{ws-ijmpa-FRK}   

\begin{document}

\markboth{G. E. Volovik}
{Vacuum Energy: Myths and Reality}

%
\catchline{}{}{}{}{}
%

\title{Vacuum Energy: Myths and Reality}

\author{G. E. VOLOVIK}

\address{Helsinki University of Technology,
Low Temperature Laboratory,\\ P.O. Box 2200, FIN--02015 HUT,  Finland, \\
Landau Institute for Theoretical Physics, 119334 Moscow, Russia\\
volovik@boojum.hut.fi}

\maketitle

\pub{Received (Day Month Year)}{}

\begin{abstract}
We discuss the main myths related to the vacuum energy and cosmological constant, such
as: ``unbearable lightness of space-time''; the dominating contribution of zero point
energy of quantum fields to the vacuum energy; non-zero vacuum energy of the false
vacuum; dependence of the vacuum energy on the overall shift of energy; the absolute
value of energy only has significance for gravity;  the vacuum energy depends on the
vacuum content; cosmological constant changes after the phase transition; 
zero-point energy of the vacuum between the plates in Casimir effect must gravitate,
that is why the zero-point energy in the vacuum outside the plates must also gravitate;
etc. All these and some other conjectures appear to be wrong when one considers the
thermodynamics of the ground state of the quantum many-body system, which mimics
macroscopic thermodynamics of quantum vacuum.  In particular, in 
spite of the ultraviolet divergence of the zero-point energy,   the natural value of
the vacuum energy is comparable with the observed dark energy. That is why the vacuum
energy is the plausible candidate for the dark energy.

\end{abstract}

\section{Introduction: Old and new myths}

Quantum mechanics killed the old myth that the vacuum is an empty space. Now 
everybody agrees that the vacuum is filled with zero-point fluctuations
of relativistic quantum fields (see reviews \cite{Weinberg,Nobbenhuis,Padmanabhan}).
Hovever, this new point created new myths.

The main myth is referred to as the main cosmological constant problem or  
the ``unbearable lightness of space-time'' \cite{Wilczek}. 
It is based on assumption 
that the natural  value of the energy density of the
quantum vacuum and thus of the cosmological constant
is determined by the (microscopic) Planck energy scale:
\begin{equation}    
 \Lambda_{\rm natural}\sim  \frac{E_{\rm Planck}^4}{\hbar^3c^3}
~,
\label{LambdaMicro}
\end{equation}
This value is fantastically too big  
compared to the observational upper limit for 
the cosmological constant, which forces us to think that our Universe is dramatically
unnatural \cite{Carroll}. 

The related myth is that the zero-point energy of bosonic quantum fields and 
the negative energy
of Dirac sea of the fermionic quantum fields are the 
dominating sources of vacuum energy. The summation over all the modes
till the  Planck energy scale gives the estimate in Eq.(\ref{LambdaMicro}).
The possible supersymmetry between bosons and fermion may reduce the cut-off, but
the result is still unbearably huge.

But does the equation (\ref{LambdaMicro})  really represent  the natural 
value of the vacuum energy?
In Sec. \ref{lightness} we find that the natural value of the vacuum energy is
determined by macroscopic physics, and it
is  by many orders of magnitude smaller than it follows from the naive summation  of
modes. And in Sec. \ref{ZeroPoint} we discuss how the zero-point energy 
and/or the negative energy of Dirac vacuum are compensated by the microscopic
(trans-Planckian) degrees of freedom without any fine-tuning.   

The discussion is based on 
our knowledge of the many-body systems,
where the ground state mimics 
the quantum vacuum in many respects, and even  the similar problems  related to  the
vacuum energy also arise. We know the microscopic (atomic=trans-Planckian)  structure
of the vacuum in the many-body system, and  thus are able to see how these problems
are solved there. We find that the vacuum energy is the macroscopic 
characteristic of the quantum vacuum, and 
its behavior is generic and  does not depend on 
the details of the microscopic physics. This 
encourages us to extend the found properties of 
the vacuum energy to the quantum vacuum of our 
Universe whose microscopic structure still remains unknown, and unveil
different  myths related to the vacuum energy.

One of the myths is that the magnitude of the vacuum energy
is only important when the gravity is present, otherwise it can be 
removed by the shift of the energy. This is not so, the consideration of the
many-body system in Sec. \ref{gravity} demonstrates that the properly determined
vacuum energy density is well defined in any system
(relativistic or non-relativistic, with gravity or without gravity)
and it does not depend on the choice of zero  (Sec. \ref{shift}). 
The macroscopic properties of the quantum vacuum 
do not depend on whether the  gravity emerges in the system or not.  
If so, the problems related to the vacuum energy
can be considered in the more general context
without only referring to the systems with gravity. 

Of course, there exist the many-body systems 
where the dynamic metric field emerges  as 
one of the collective modes of the  quantum vacuum.
Moreover, we have examples where gravity emerges together with 
the ingredients of the  Standard Model -- chiral fermions and gauge fields
\cite{VolovikBook}.
These are the Fermi systems with the topologically stable Fermi points 
in the fermionic spectrum, where the relativistic quantum fields
and gravity emerge due to
the so-called 
Atiyah-Bott-Shapiro construction \cite{Horava}.
However, this  encouraging fact of the existence 
of the class of many-body systems which almost 
perfectly mimic the vacuum of the Standard Model  
is not important for our discussion
of the more general properties of the quantum vacuum related
to vacuum energy.

The next myth is about the energy of the false vacuum.  The false  vacuum
is  the state corresponding to the local minimum of energy. It has higher
energy   than the true vacuum which corresponds to 
absolute minimum. That is why, it is always assumed that 
if $\Lambda=0$ in the true vacuum, it must be positive 
and big in the false vacuum.
This is important for the phenomenon of inflation -- 
the exponential super-luminal expansion of the
Universe.  
The related myth is the effect of the symmetry breaking phase transition,
such as  electroweak phase transition or chiral phase transition in
strong interactions,  on the vacuum energy. Both lead to extremely large
values of vacuum energy relative to what is allowed. In Sec. \ref{FalseVac} we discuss
how the condensed matter systems treat these problems.

The other related myth is that the vacuum energy depends on the vacuum content. 
Indeed, at first glance this is correct: the vacuum is complicated, there are 
many contributions 
to vacuum energy from different quantum fields. There are 
many species of fermions each with its own mass. Thus
the vacuum energy must depend on the fermion masses, 
on Higgs field, etc. (see e.g.
\cite{Polchinski}). We discuss this problem in Sec. \ref{VacuumContent}. 

As follows from the condensed matter experience, the short answer to the above 
problems is that the natural value of the vacuum energy is zero, and thus does not
depend on the vacuum content, on the divergence of the zero-point energy, on whether 
the vacuum is false or true, on the phase transition in the vacuum, etc. 

If the natural value of the vacuum energy is zero should  we return to the old  myth
that the vacuum does not gravitate? We should not, since the non-zero cosmological
constant has been observed producing again the new myth that this observation  has
catalyzed the crisis in physics \cite{Polchinski}. It is easier to assume  that the
vacuum energy is zero than to explain the reduction by 120  orders of magnitude. We
discuss this point in Sec. \ref{NonGrVac} and demonstrate how the deviations of the
vacuum from the perfectness naturally induce the non-zero   vacuum energy comparable
with observations.

One of the perturbations of the vacuum which lead to the non-zero vacuum energy  is
the boundary conditions. This perturbation  produces the Casimir effect. In Sec.
\ref{Lamb} we discuss: why the zero-point energy  between the plates in the Casimir
effect must gravitate, while the zero-point energy in the vacuum outside the plates
does not gravitate; and the related question posed in Ref. \cite{Polchinski} --  why
the zero-point energy must gravitate in the environment of the atom  (Lamb shift) and
not in vacuum. 

\section{What is natural value of vacuum energy?}
\label{lightness}

Let us start with the main cosmological constant problem, which
is based on the myth  that the natural  value of the energy density of the
quantum vacuum and thus of the cosmological constant
is determined by the Planck energy scale. So, what is the natural 
value of the vacuum energy?

\subsection{Natural microscopic value}
\label{NaturalMicroscopic}

To get some insight into this problem, 
let us consider whether this assumption true or not in
quantum many-body systems, 
such as quantum liquid or solid. At first glance this
is true. The microscopic (atomic) physics determines the characteristic
scales for the quantities describing liquids and 
solids.
The role of the Planck length scale is played 
by the inter-atomic distance $a$. Correspondingly the 
role of the Planck energy scale is played either by the
Debye temperature,  or by melting temperature, 
or by the transition temperature to the superfluid state.
As a result we have the following estimates for the characteristic temperature,
energy density and pressure: 
\begin{equation}    
T_{\rm micro} \sim E_{\rm Planck}\sim \hbar c/a~~,
~~ \epsilon_{\rm micro}\sim p_{\rm micro} 
\sim \frac{E_{\rm Planck}^4}{\hbar^3c^3}~.
\label{CharacterTP}
\end{equation}
Here $c$ is the speed of sound waves or 
of other relevant bosonic or fermionic excitations, which
plays the role of the maximum attainable speed (the speed of light).

\subsection{Natural macroscopic value}
\label{NaturalMacroscopic}

In spite of the similarity between Eq.(\ref{LambdaMicro}) and 
Eq.(\ref{CharacterTP}), the conjecture that 
the microscopic (Planck) scales naturally enter the vacuum energy 
is not correct.
This is because the real temperature and real pressure of 
liquids and solids are macroscopic variables, 
which are determined not by the internal microscopic physics but by the environment.
In particular, if there is no interaction with the environment, 
the many-body system will be cooled down to zero temperature
by radiation and/or evaporation, 
while its pressure in the final equilibrium state will be zero, 
because the external pressure is absent:
\begin{equation}    
T_{\rm macro}=0~~,~~ p_{\rm macro} =0~.
\label{RealTP}
\end{equation}
This demonstrates the huge difference between the 
parameters of the microscopic and macroscopic physics. 
The many-body system as a whole obeys the macroscopic physics
which does not depend on the details of the microscopic physics.
Since this is the  general property of any system,
it would be strange if the vacuum does not obey the same rule. 
The vacuum pressure  belongs to the macroscopic physics, 
and thus  the natural value of the vacuum energy density must be zero:
\begin{equation}    
 \Lambda_{\rm natural}=\epsilon_{\rm vac}=-p_{\rm vac}=0
~,
\label{LambdaMacro}
\end{equation}
if we believe that our Universe does not interact with the environment.

\section{The absolute value of energy only has significance for
gravity}
\label{gravity}

 One may argue that the vacuum energy is only important for gravity, 
i.e. only gravity is sensitive to vacuum energy, 
otherwise the vacuum energy can be removed by the shift of the zero level. 
That is why the 
analogy with condensed matter makes no sense 
(unless gravity emerges in condensed matter).  This is another myth.

First of all, we are interested not in the absolute value of the total
vacuum energy. We are interested in the cosmological constant, which
represents the density of the vacuum energy, and this density is related
to the vacuum pressure by the equation of state
\begin{equation}    
\epsilon_{\rm vac}=-p_{\rm vac} 
~,
\label{VacEOS}
\end{equation}
Irrespective of whether the gravity is present or absent,  the vacuum is
the substance with this equation of state, and this does not depend on the
choice of the zero.

Let us consider how the equation of state (\ref{VacEOS}) 
emerges for the vacuum   (ground state) of condensed matter.

\subsection{Quantum field theory of  condensed matter}
\label{QFTcondmat}

The rules of the game are simple: the ground state of condensed matter
is the analogue of the vacuum. Particle-like excitations above the ground state -- quasiparticles --
serve as analogue of matter:  quasiparticles do not scatter on the underlying atoms of the liquid or solid if the  liquid or solid is in its ground state. That is why for quasiparticles, the homogeneous ground state is the perfect vacuum.

First, one must specify what thermodynamic potential in the
quantum condensed matter plays 
the role of the vacuum energy density. 
Let us start with the "Theory of Everything" for condensed matter
system. The underlying microscopic physics of a liquid or solid formed by 
a system of $N$ atoms obeys the conventional quantum mechanics  
and is described by the $N$-body Schr\"odinger wave function  $\Psi({\bf
r}_1,{\bf r}_2,
\ldots  , {\bf r}_i,\ldots  ,{\bf r}_N)$. 
The corresponding many-body Hamiltonian is
\begin{equation}
{\cal H}= -{\hbar^2\over 2m}\sum_{i=1}^N {\partial^2\over \partial{\bf r}_i^2}
+\sum_{i=1}^N\sum_{j=i+1}^N U({\bf r}_i-{\bf r}_j)~,
\label{TheoryOfEverythingOrdinary}
\end{equation}
where $m$ is the bare
mass of the atom, and  $U({\bf r}_i-{\bf r}_j)$ is the
pair  interaction of the bare atoms $i$ and $j$.

In the thermodynamic limit where the
volume of the system $V\rightarrow
\infty$ and $N$ is macroscopically large, there emerges an equivalent
description of the system in terms of quantum fields, in a procedure
 known as second quantization.  
The quantum field is presented by the bosonic 
or fermionic annihilation operator for atoms
$\psi({\bf x})$. 
The Schr\"odinger many-body Hamiltonian
(\ref{TheoryOfEverythingOrdinary}) is transformed to the Hamiltonian of the
quantum field theory (QFT) \cite{AGDbook}:
\begin{eqnarray}
\hat H_{\rm QFT}=\hat H-\mu \hat N=\int d{\bf x}\psi^\dagger({\bf
x})\left[-{\nabla^2\over 2m} -\mu
\right]\psi({\bf x}) \nonumber \\+{1\over 2}\int d{\bf x}d{\bf y}U({\bf
x}-{\bf y})\psi^\dagger({\bf x})
\psi^\dagger({\bf y})\psi({\bf y})\psi({\bf x}).
\label{TheoryOfEverything}
\end{eqnarray}
Here $\hat N=\int d^3x~\psi^\dagger({\bf
x})\psi({\bf x})$ is the operator of the particle number (number of
atoms); $\mu$ is the chemical potential 
-- the Lagrange multiplier introduced to take into account the
conservation of the number of atoms.

One can see that in condensed matter, the emerging quantum field theory 
is governed not by the Hamiltonian $\hat H$ but 
by the Hamiltonian  $\hat H_{\rm QFT}=\hat H-\mu \hat N$ which allows us
to avoid the constraint imposed on the quantum field $\psi$ by the
conservation of particle number. The Hamiltonian
(\ref{TheoryOfEverything}) serves as a starting point for the
construction of the effective QFT for quasiparticles living at low
energy, and it is responsible for their vacuum.  In the  QFT description,
the energy density of the vacuum is  given by the thermodynamic potential
$E-\mu N$,  which is the vacuum expectation value of the Hamiltonian
$\hat H_{\rm QFT}$:
\begin{equation}
\epsilon_{\rm vac}  =\frac{1}{V}\left<\hat H_{\rm QFT}\right>_{\rm vac}~.
\label{VacuumEnergy}
\end{equation}

\subsection{Equation of state for vacuum in condensed matter}
\label{EoScondmat}

 One can also check that the energy density (\ref{VacuumEnergy}) 
is the right choice for
the vacuum energy density by using the Gibbs-Duhem relation of
thermodynamics, which is valid  in the thermodynamic 
limit $N\rightarrow \infty$. It states that if the condensed matter 
is in equilibrium it obeys the following relation between the energy
$E=\left<{\hat H}\right>$, and  the other thermodynamic variables --
the temperature  $T$, the entropy
$S$, the particle numbers $N=\left<{\hat N}\right>$, the
chemical potentials $\mu$, and the pressure $p$: 
\begin{equation}
E-TS-  \mu  N  =-pV~.
\label{Gibbs-DuhemRelation}
\end{equation}
Applying this relation to the equilibrium vacuum, 
i.e. to the equilibrium state at $T=0$, 
one obtains the relation (\ref{VacEOS}) between the  vacuum pressure and  
vacuum energy
density:
\begin{equation}
\epsilon_{\rm vac}  =\frac{1}{V}\left<\hat H_{\rm QFT}\right>_{\rm vac}
=\frac{1}{V}\left<\hat H\right>_{\rm vac}-
\frac{1}{V}\mu \left<\hat N\right>_{\rm vac} =-p_{\rm vac}~.
\label{Gibbs-DuhemEoS}
\end{equation}

\subsection{Thermodynamics of vacuum}
\label{thermodynamics}

\begin{figure}
  \includegraphics[height=0.4\textheight]{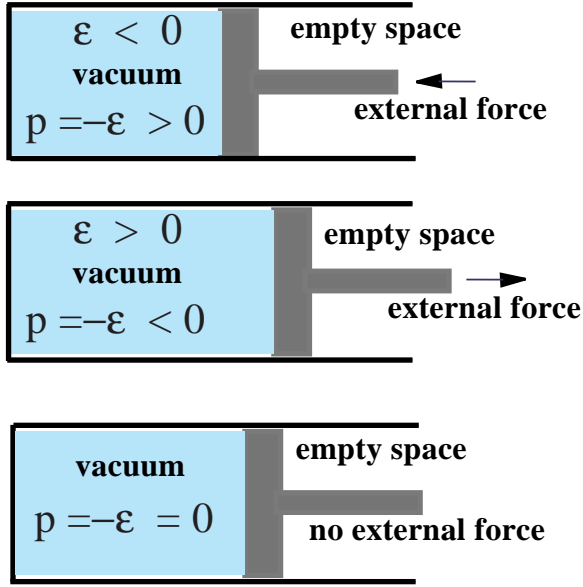}
  \caption{
If the vacuum energy is negative, the vacuum tries to expand 
by moving the piston to the right. To reach an equilibrium, the external
force must be applied which pushes the piston to the left and compensates for
the positive  vacuum pressure, $p_{\rm vac}=-\epsilon_{\rm vac}$. 
The same takes place in condensed matter --  gases, liquids
and solids -- at $T=0$.
In the same manner, if the vacuum
energy is positive, the applied force  must pull the piston to the
right to compensate for the negative vacuum pressure. The same takes place in
 liquids and solids at $T=0$, but not in  gases which can be in equilibrium only 
 at positive pressure. As distinct from gases, 
the liquids and solids are self-sustained
substances: they can be in equilibrium in the absence of external forces. 
Thus in the absence of interaction with environment, 
the energy density of the vacuum and vacuum pressure acquire
their natural values, $\epsilon_{\rm vac}=-p_{\rm vac}=0$.
 }
\label{Piston}
\end{figure}

This thermodynamic analysis does not depend on the microscopic 
structure of the vacuum and thus can be applied to any quantum vacuum. 
It follows from the simple analysis 
of the equilibrium of the piston between the quantum vacuum and 
the empty space
(see Fig. \ref{Piston}). The force acting on the piston is 
$F=-dE_{\rm vac}/dx=-AdE_{\rm vac}/dV=-A\epsilon_{\rm vac}$, 
where $A$ is the area of the piston. On the other hand the force divided 
by area $A$ must be equal to the vacuum pressure. 
This gives the equation of state
(\ref{VacEOS})  for the vacuum.

It is the general property, which follows from thermodynamics, that the
vacuum behaves as a medium with the equation of state (\ref{VacEOS}). 
Thus it is
not surprising that the equation of state is applicable
also to the particular case of the vacuum of the relativistic quantum 
field theory (RQFT). This demonstrates that the problem of the vacuum energy
can be considered from a more general perspective not constrained by
the relativistic Hamiltonians. Moreover,  it is not important whether
gravity emerges or not in the system, i.e. whether the vacuum is gravitating 
or not.

The vacuum energy plays an important role even in the absence of 
gravity \cite{VacSpecRel}. Let us imagine the world without gravity, 
i.e. the world where the metric field does not emerge
in the low-energy corner, or the world where the Newton constant $G=0$. In such a world
the matter would expand due to the matter pressure. However, comparing this situation with
what occurs in condensed matter, where the role of the vacuum is played by the ground
state, and the role of matter  is played by quasiparticles, we shall see in the Sec.
\ref{ResponseTomatter} that the Universe in such world can be stabilized by the vacuum
pressure which compensates the pressure of matter.

\subsection{What is $\mu$ in quantum vacuum?}
\label{mu}

The frequently asked question is what is the analog of the chemical 
potential $\mu$ in the underlying microscopic physics of the quantum vacuum 
in our Universe. In all the known quantum condensed matter system one has 
one or several conservation laws for different atoms or even for different 
atomic states of the same atoms.  The general Hamiltonian
for $l$ quantum fields $\psi_i$ ($i=1,...,l$) for $l$ species of atoms is
\begin{equation}
\hat H_{\rm QFT}=\hat H-\sum_{i=1}^l\mu_i \hat N_i~.
\label{ManyAtoms}
\end{equation}
Application of the Gibbs-Duhem relation to this system,
\begin{equation}
E-TS-\sum_{i=1}^l\mu_i  N_i=-pV~,
\label{GDManyAtoms}
\end{equation}
again gives the equation of state for the quantum vacuum (the equilibrium state at $T=0$)
\begin{equation}
\epsilon_{\rm vac}  =
\frac{1}{V}\left<\hat H_{\rm QFT}\right>_{\rm vac}=-p_{\rm vac}~.
\label{ManyAtomsGibbs-DuhemEoS}
\end{equation}
This equation of state does not depend on the number of the conservation laws, 
which means that the underlying physics of quantum vacuum may have any number 
$l$  of chemical potentials, including $l=0$. 
If for our Universe one has $l=0$, then the question
asked in the title of this sub-section makes no sense. 

However, there is an opinion that, in order for the condensed-matter approach to work
for the quantum vacuum, there should exist at least one chemical potential, and it is
the challenge of the approach to find out its origin
\cite{Bjorken}. This is true that among the quantum condensed matter systems we do not
have examples of a stable system with $l=0$. But this does
not mean that the quantum vacuum of our Universe must have nonzero
$l$: there is no reason for the structure of the quantum vacuum to
mimic exactly the condensed matter.  In principle, the non-zero $l$ is not excluded,
and the candidate for the conserved quantity can be, for example, the fermion number.
In Appendix we use the model of the quantum vacuum with $l=1$ for the discussion of
the cosmological constant emerging in Einstein Universe.

\section{Does vacuum energy depend on overall shift?}
\label{shift}

 Now we can turn to another myth, that the vacuum energy depends on 
the overall energy shift.

This is not so, because the vacuum pressure is the intensive 
thermodynamic variable. It is invariant under  the overall energy shift: 
under  transformation $E\rightarrow E+E_0$  the pressure does not change, 
$p=-dE/dV \rightarrow p$. The vacuum energy density is also the intensive 
thermodynamic variable, and thus it must be also invariant  under  
the overall energy shift. This is seen from the equation of state 
(\ref{VacEOS}) for the vacuum, which relates the energy density to pressure.

The other frequent comment related to the shift of the energy is that  
the energy per particle in Eq.(\ref{TheoryOfEverythingOrdinary}) 
for condensed matter  is not defined in unique way. For example, one 
may add the rest energy $mc^2$ to the energy of each particle, 
$p^2/2m \rightarrow mc^2 +p^2/m$ or add the constant potential 
$p^2/2m \rightarrow  p^2/m +U_0$. This leads to the overall energy shift with
$E_0=Nmc^2$ or $E_0=NU_0$  which now depends on the volume $V$. 
However, the thermodynamic quantities do not depend on such transformation. 
In particular, the properly defined vacuum energy density in 
Eq.(\ref{VacuumEnergy}) is invariant under this transformation, 
since the corresponding transformation of the chemical potential, 
$\mu\rightarrow \mu+mc^2$ or $\mu\rightarrow \mu+U_0$ correspondingly, 
compensates the change in $E$ leaving $\epsilon_{\rm vac}$ invariant.

\section{Is zero-point energy dominating in vacuum energy? }
\label{ZeroPoint}

\subsection{Zero-point contribution}
\label{NaiveEstimation}
 
From Sec. \ref{NaturalMacroscopic} it follows  that the natural value of the vacuum
pressure of the system in  the absence of the environment is $p_{\rm vac}=0$. 
Then from the equation of state for the vacuum
(\ref{VacEOS})  it follows that the natural value of the vacuum energy 
density must be also zero. But this poses the  problem:  
what to do with the zero-point energy,
The zero value of the  vacuum energy is in huge contradiction 
with our next myth:
the  positive zero-point energy of bosonic quantum fields and 
the negative energy
of Dirac vacuum of the fermionic quantum fields are the 
dominating sources of vacuum energy, 
each comprising the Planck scale estimate (\ref{LambdaMicro}) 
for the vacuum energy and cosmological constant.
The naive summation over the $\nu_b$ bosonic and $\nu_f$ 
fermionic modes of the
quantum fields gives
\begin{equation}
\epsilon_{\rm vac}= {1\over V}\left(  {1\over 2} \sum_b \sum_{\bf
p}  cp~~
- \sum_f
\sum_{\bf p} cp\right)\sim {1\over c^3} \left({1\over 2}\nu_b 
-\nu_f\right) E_{\rm Planck}^4~.
\label{VacuumEnergyPlanck1}
\end{equation}

This contradiction between the  zero value, which follows from 
the macroscopic physics,  and the Planck scale value following 
from the microscopic (Planckian) physics exists also in the many-body system.
Observers  living in such a system would also make the same estimation, 
which is based on their knowledge of the low-energy bosonic and fermionic 
fields with corresponding parameters $c$, $E_{\rm Planck}$, 
$\nu_f$ and $\nu_b$ appropriate for a given system. 
For example, if they live in solids, their life is based on the 
``relativistic'' bosonic fields of 
phonons. They are aware that there is the Planck energy scale 
(played by Debye temperature) where the ``Lorentz invariance'' is violated, i.e.
the spectrum of phonons starts to deviate from the linear ``relativistic'' spectrum. 
That is why these inner observers will be able to estimate theoretically
the vacuum energy as zero-point energy of the phonon field, and they will be 
also puzzled by the disparity of many orders of magnitude between the
estimates and observations.

 \subsection{Automatic compensation of zero-point energy}
\label{CompensNaiveEstimation}

But we know the microscopic physics of solids and liquids, and we can easily
explain to the inner observers where their theory goes wrong.  
Actually there is nothing wrong with their estimation, it is simply incomplete.
The equation (\ref{VacuumEnergyPlanck1}) takes into account only the modes below the ``Planck'' energy cut-off, i.e. 
those degrees of freedom which are described by
an effective theory (theory of elasticity in solids or hydrodynamics in liquids). At higher energies, 
which correspond to distances of order of inter-atomic spacing $a$,
the microscopic structure of solids or liquids in terms of the interacting
atoms in Eq.~(\ref{TheoryOfEverything}) must be taken into account. They 
 cannot be expressed in terms of the relativistic quantum fields 
 of phonons. 
When one sums up all the contributions to the vacuum energy, 
sub-Planckian (phonons) and  trans-Planckian (all other degrees of freedom), 
one obtains the zero result (microscopic calculations of the ground state energy of
the system of $N$ atoms obeying the `Theory of 
Everything in Eq.(\ref{TheoryOfEverythingOrdinary}) can be found in Ref.
\cite{Microscopic}; one can check that $\left<\hat H\right>_{\rm vac}=
 \mu \left<\hat N\right>_{\rm vac}$ and thus $\epsilon_{\rm vac}=0$). The exact
nullification occurs without any special fine-tuning,  it is provided by the
macroscopic physics -- thermodynamics -- applied to 
the whole equilibrium vacuum.

The main lesson from us to inner observers (or from condensed matter
to particle physics), which they may or may
not accept, is this: 
the energy density of the homogeneous equilibrium state of the
quantum vacuum is zero in the absence of an external environment. The
higher-energy  (trans-Planckian, microscopic) degrees of freedom of the 
quantum vacuum,
whatever they are,  perfectly cancel the huge positive contribution of the
zero-point fluctuations of the quantum fields as well as the huge negative
contribution of the Dirac vacuum. 
This is the consequence of the macroscopic physics which forces 
the microscopic degrees of freedom to automatically fine-tune.

This effect of the automatic compensation without fine tuning can be found
also in some relativistic theories. Example is provided by a  model, 
in which our world is represented by the $(3+1)$-dimensional membrane embedded in 
the $(4+1)$-dimensional anti-de Sitter space.  
In the  equilibrium vacuum, the  huge contributions to the  cosmological constant
coming from different sources cancel each other without fine-tuning \cite{Andrianov}.

 \subsection{Myth in condensed matter community}
\label{MythCondMat}

It happens, that the myth on the zero-point energy
is still alive even in condensed matter
community.
Some people erroneously add the zero-point energy of the quantized
phonon field to the energy of the liquid or solid.
This leads, however, to the double counting. Phonons are
quanta of the classical sound waves propagating on the background 
of the liquid or solid. Their spectrum is $E(p)=cp$, where the 
speed of sound $c$ is determined by the energy density of the system as a function of the
mass density: $c^2=\rho d^2\epsilon/d\rho^2$.  
The energy $\epsilon(\rho)$ of the background on which the phonons 
are propagating takes into account already all the quantum degrees
of freedom of underlying liquid, including those which
can be expressed via phonons --  the quantized field of small oscillations
of liquid or solid. 

In other words, sound waves represent 
the classical output of the quantum system,
in which phonons are already quantized from the very beginning.
By quantizing phonons again one only reproduces 
the low-energy part of already existing quantum states of the system.
That is why one makes mistake if one adds the zero-point energy
of phonons to the overall ground state energy of the system.

\section{Does vacuum energy depend on vacuum content?}
\label{VacuumContent}

The next myth is that the vacuum energy depends on the vacuum content. 
Indeed, at first glance this is correct: the vacuum is complicated, there are 
many contributions 
to vacuum energy from different quantum fields \cite{Polchinski}. There are 
many species of fermions each with its own mass. Thus
the vacuum energy must depend on the fermion masses, 
on Higgs field, etc. (see e.g.
\cite{Polchinski}). For example, the  
general form of the contribution
to the vacuum energy density from the electron-positron degrees of 
freedom of the quantum vacuum is
\begin{equation}
\epsilon  =a_4 M^4 + a_2M^2m^2 + a_0m^4\ln{M^2\over m^2}~,
\label{General}
\end{equation} 
where $m$ is the electron mass, and $M$ is the
ultraviolet energy cut-off. If the cut-off is provided by the Planck scale, 
the estimate (\ref{LambdaMicro}) is obtained (we use units $\hbar=c=1$).

Different regularization schemes were
suggested in order to obtain the dimensionless parameters $a_i$.
However, this is not relevant for the estimation of the total vacuum 
energy density. From condensed matter experience we know that  condensed
matter may contain  different species of atoms and zero-point energy of
different effective bosonic fields (phonons, magnons, etc.).
Nevertheless, the vacuum energy density relevant for quantum field theory is
the macroscopic parameter: it does not depend on the microscopic physics and thus on the vacuum content. It remains zero in a full equilibrium.

The particular contributions to the vacuum energy becomes important, when we consider the 
coexistence of two vacua with slightly different vacuum content, 
for example with the slightly different $m$ \cite{PomeranchukProc}.

\section{Myth on energy of false vacuum }
\label{FalseVac}

Let us turn to the myth on the energy of the false vacuum.  The false  vacuum
corresponds to the local minimum of energy, and it has higher energy than the true
vacuum which corresponds to  absolute minimum. That is why,  
if $\Lambda=0$ in the true vacuum (as is typically assumed), it must be nonzero
(positive) in the false vacuum; or the other way round: 
if $\Lambda=0$ in a false vacuum, it must be nonzero (negative) in the true vacuum.

This is important for the phenomenon of inflation -- 
the exponential super-luminal expansion of the
Universe. In some theories, the inflation is caused by a false vacuum. It
is usually assumed that the energy of the true vacuum is zero, and thus
the energy of the false vacuum must be positive. Though the false vacuum
can be locally stable at the beginning,   $\Lambda$  in this vacuum must
be  big and positive constant, which causes the exponential de-Sitter
expansion. This scenario is wrong, because as we know the cocally stable quantum vacuum does not gravitate.

The related myth is the effect of the symmetry breaking phase transition,
such as  electroweak phase transition or chiral phase transition in
strong interactions,  on the vacuum energy. Both lead to extremely large
values of vacuum energy relative to what is allowed.

The related question also, why the cosmological constant is (approximately) zero only
for that vacuum in which we leave? Our vacuum has no special preference compared to
the other possible minima of the effective potential. To avoid this problem, the idea
of the Multiple Point Principle has been introduced, which states that the
vacuum energy must be (approximately) zero in all of the vacua 
\cite{FrogattNielsen2003}. 

Let us look at all these problems related to vacua corresponding to local minima using our
knowledge of the general thermodynamic properties of the quantum vacuum. 

\subsection{Cosmological constant in false and true vacua}
\label{CCFalseVac}

Analyzing the Gibbs-Duhem relation we find that in our derivation of 
the vacuum energy, we never use the fact that our system is in the true
ground state. We only assume that our system is in the
thermodynamic equilibrium at $T=0$. But this is applicable to 
the metastable state
too if we neglect the tiny, exponentially small probability of the  transition between
the false and true vacua, such as quantum tunneling  and thermal activation. Thus we
come to the following, at first glance paradoxical, conclusion:
the cosmological constant in all  homogeneous vacua in equilibrium is zero,
irrespective of whether the vacuum is true or false. This poses
constraints on some scenarios of inflation.

\begin{figure} 
\includegraphics[width=\textwidth]{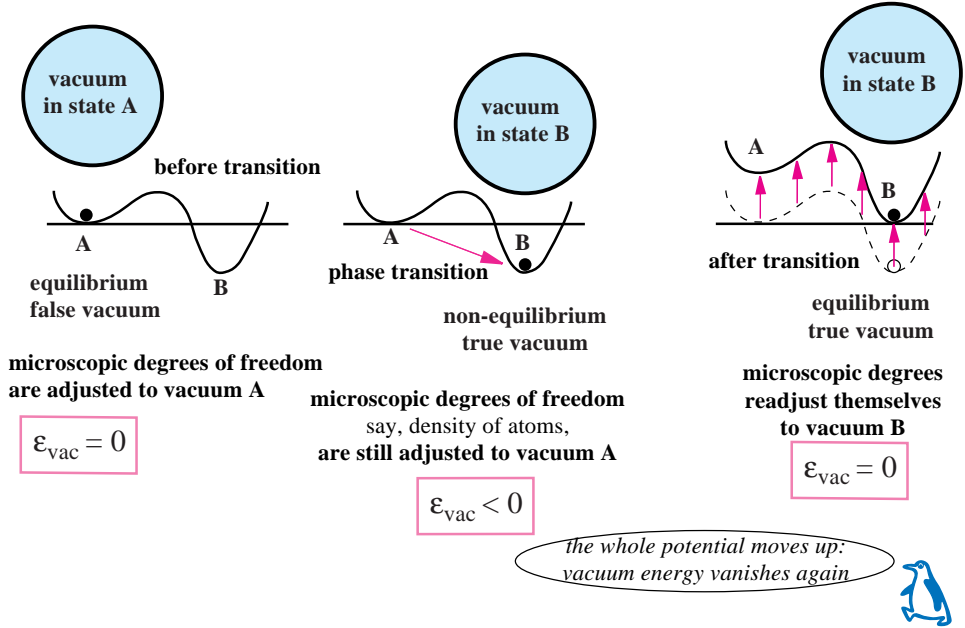}
\caption{ The condensed-matter scenario of the evolution of the energy
density
$\epsilon_{\rm vac}$ of the quantum vacuum in the process of  the first
order phase transition from the equilibrium false vacuum to the
equilibrium true vacuum. Before the phase
transition, i.e. in the false but equilibrium vacuum, one has 
$\epsilon_{\rm vac}=0$. During the transient period the microscopic
parameters of the vacuum readjust themselves to new equilibrium state,
where the equilibrium condition 
$\epsilon_{\rm vac}=0$ is restored.}
\label{transition}
\end{figure}

\subsection{How the phase transition occurs}
\label{PT}

The related myth is the effect of the symmetry breaking phase transition
on the vacuum energy. One may ask, how and why the transition between the
vacua occurs  if they have the  same energy density. Again the answer
comes from the analysis of the  phase transition in condensed matter
\cite{VolovikAnnals}. 

Let us consider
the typical example of the first-order  phase transition which occurs 
between the metastable quantum liquid $^3$He-A and the stable quantum
liquid  $^3$He-B, Fig. \ref{transition}.  Initially the vacuum is in the state A.
Since this state is the local minimum, its thermodynamic potential relevant for the 
vacuum energy is zero,  $E_A-\mu_A N=0$.  Let us assume that this vacuum A  is metastable, i.e.
there exists the vacuum B for which the energy calculated at the same fixed chemical potential $\mu=\mu_A$ is negative,
$E_B-\mu_A N<0$. Thus the liquid prefers the phase transition from
the ``false'' vacuum  A to the ``true'' vacuum  B. When the transition to the B-phase
occurs, the vacuum energy becomes negative. However, this corresponds to the
non-equilibrium state, since we know that in equilibrium the vacuum energy must be zero. 
What happens? The microscopic degrees of freedom start to rearrange themselves to the
new vacuum B. On the microscopic level this means that the
inter-atomic spacing changes in such a way that the chemical potential $\mu_B$ in the new vacuum
starts to satisfy the equilibrium condition, $E_B-\mu_B N=0$. This means that after
some transient period of relaxation the vacuum energy density $\epsilon_{\rm vac}$ in the true vacuum B also becomes zero.

We can readily apply this consideration to the quantum vacuum 
in our Universe. This condensed-matter example suggests that the
cosmological constant is zero before the cosmological phase transition.
During the non-equilibrium transient period of time, the microscopic
(Planckian) parameters of our vacuum are adjusted to a new equilibrium
state in a new vacuum, and after that the cosmological constant becomes
zero again.  Thus vacuum energy in each vacuum is zero, but this only occurs when the
system is in  equilibrium within a given vacuum.  This demonstrates that to have the
zero cosmological constant in each vacuum we do not need the Multiple Point Principle
suggested in Ref.
\cite{FrogattNielsen2003}, which states that the effective potential in Fig.
\ref{transition} should have degenerate minima. This does not happen in condensed matter: minima are typically
non-degenerate (if there is no special symmetry). They will be degenerate only in the
special case when two (or several) vacua coexist in space across the interface
\cite{MPP}.

We do not know what are the  microscopic
degrees of freedom in our Universe and how the microscopic parameters relax in the new vacuum to establish the new
equilibrium after phase transition. This already depends on the details of the system
and cannot be extracted from the analogy with quantum vacua in liquids. However,
using our experience with quantum liquids we can try to estimate the
range of change of the  microscopic parameters after the
transition.

\subsection{Vacuum response to electroweak phase transition}
\label{Electroweak}

Let us consider, for example, the electroweak phase transition, 
assuming that it is of the first order and thus can occur at low
temperature, so that we can discuss the transition in terms of the vacuum
energy. In this transition, the vacuum energy density changes from zero
in the initially equilibrium false vacuum to the negative value on the
order of 
\begin{equation}
\delta \epsilon_{\rm vac}^{\rm ew}\sim -E_{\rm ew}^4 
\label{ElectroweakCorrection}
\end{equation}
in the true vacuum, where $E_{\rm ew}$ is the electroweak energy scale. 
 To restore the equilibrium in the new vacuum, this negative energy must be compensated by
the adjustment of the microscopic (trans-Planckian) parameters. As such a
parameter we can use the value of Planck energy $E_{\rm Planck}$, since  
it determines the characteristic microscopic energy scale.
We know that the contribution of the modes close to or above the Planck energy to
the vacuum energy is huge, since in equilibrium it must compensate the $E_{\rm
Planck}^4$ contribution from the sub-Planckian modes.
The variation of this microscopic energy due to the change of
the parameter $E_{\rm Planck}$ is
\begin{equation}
\delta \epsilon_{\rm vac}^{\rm Planck} 
\sim E_{\rm Planck}^3 \delta E_{\rm Planck}~.
\label{Correction}
\end{equation}
In a new equilibrium vacuum, the density of the vacuum energy must be
zero   
\begin{equation}
 \delta\epsilon_{\rm vac}^{\rm ew} +\delta\epsilon_{\rm vac}^{\rm
Planck} =0~,
\label{Compensation}
\end{equation}
and thus the relative change of the microscopic parameter
$E_{\rm Planck}$ which compensates the change of the electroweak energy
after the transition is
\begin{equation}
\frac{\delta E_{\rm Planck}}{E_{\rm Planck}} \sim 
\frac{E_{\rm ew}^4}{E_{\rm Planck}^4}~.
\label{Correction2}
\end{equation}
This, for example, leads to the tiny change of the Newton constant, which is
determined by the Planck scale, $G\sim 1/E_{\rm Planck}^2$:
\begin{equation}
\frac{\delta G}{G} \sim 
\frac{E_{\rm ew}^4}{E_{\rm Planck}^4}\sim 10^{-65}~.
\label{CorrectionG}
\end{equation}
The response of the deep
vacuum to the electroweak transition appears to be extremely small:  the energy at the microscopic Planck scale is
so high that a tiny variation of the microscopic parameters is enough to restore the
equilibrium violated by the cosmological transition.

The same actually occurs at the first-order
phase transition between $^3$He-A and $^3$He-B: the change in the energy 
of the superfluid  vacuum after the transition is compensated by a tiny
change of the microscopic parameter, the inter-atomic distance, $\delta a/a
\sim 10^{-6}$.

This remarkable fact of the response of the deep vacuum to  the cosmological transition may have some consequences for the dynamics of the cosmological constant after the phase transition.
Probably this would imply that  vacuum energy  relaxes rapidly, with the characteristic time determined
by the trans-Planckian physics. However,  at the moment
we have no reliable theory describing the processes of relaxation of
$\Lambda$.

\section{Myth on non-gravitating vacuum}
\label{NonGrVac}

The natural value of the vacuum energy is zero. 
Should we return to the old myth that the
vacuum does not gravitate? Of course, not, though it is easier to assume 
that the vacuum energy is zero than to explain the reduction by 120 
orders of magnitude. In particular, there exists a rather broad belief that  the problem
of the vacuum energy can be avoided simply by the proper choice of the
ordering of the QFT operators $\psi$ and $\psi^\dagger$. 
Let us consider how the quantum many-body systems help us to solve the 
cosmological constant problem $\#$2 -- why  the vacuum energy is nonzero. 

For quantum liquids or solids, the zero result has been obtained
using the original pre-QFT microscopic theory -- the Schr\"odinger
quantum mechanics of interacting atoms, from which the QFT emerges as a
secondary (second-quantized) theory.  In this approach the problem of the
ordering of the operators in the emergent QFT is resolved on the
microscopic level, and it has nothing to do with the calculations
of the vacuum energy density determined by the macroscopic physics.

The  zero result  for the vacuum energy has been obtained for the 
perfect equilibrium non-perturbed vacuum in the absence of the interaction 
with the environment.
It was assumed that the system satisfies the following conditions: 
(i) it is in complete thermodynamic equilibrium; (ii) static or stationary; 
(iii) homogeneous in space, (iv) with zero curvature; 
(v) does not contain excitations (which play the role of matter);  (vi)
does  not interact with environment; (vii) is at $T=0$; (viii) has no
boundaries.  What happens, when any of the above conditions is violated? 
Then from the same thermodynamic analysis one finds that almost all the deviations from the perfectness lead  to non-zero value of the local vacuum energy density, 
which is proportional to perturbations of the perfect vacuum.

\subsection{Vacuum response to matter}
\label{ResponseTomatter}

Let us start with the Universe without gravity, i.e. suppose that we live
in the world  obeying the laws of special relativity without general relativity. Is the static Universe
possible under this condition? Of course, it is possible if there is no matter. If matter is present, we have a problem: for $T\neq 0$ matter has positive pressure and thus will expand.
It happens that the Universe with matter can be stabilized by the pressure of the vacuum, 
and this provides an example of the importance of the vacuum energy even in the absence of gravity.

For the Universe to be static,   the pressure in the equilibrium
Universe must be zero if there is no external environment, and thus the vacuum pressure must
compensate the pressure of matter:
\begin{equation}
 p_{\rm
total}=p_{\rm matter}+ p_{\rm vac}=0 
~.
\label{ZeroPressureSpecialRelativity}
\end{equation}
 This gives the relation between the energy density of matter and the energy density of the  vacuum
 in the static world without gravity:
\begin{equation}
\epsilon_{\rm vac}=-p_{\rm vac} =p_{\rm matter}=
w_{\rm matter}\epsilon_{\rm matter}~,
\label{SpecialRelativity}
\end{equation}
where $p_{\rm matter}=w_{\rm matter}\epsilon_{\rm matter}$ 
is the equation of state for matter. The response of the vacuum energy  to matter does not depend on the details of the microscopic (trans-Planckian) physics.

Exactly the same relation
 \begin{equation}
\epsilon_{\rm vac}=
w_{\rm matter}\epsilon_{\rm matter}~,
\label{SpecialRelativityCondM}
\end{equation}
takes place in condensed matter systems. Let us consider a droplet of  a
quantum liquid or a piece of a solid in space, i.e. in the absence of the
environment, but now at non-zero temperature. At  $T\neq 0$, one has in
addition to the vacuum the gas of quasiparticles -- phonons -- which play
the role of matter. The gas of  `relativistic' phonons  has its own
partial pressure:
\begin{equation}
p_{\rm matter}=\frac{\pi^2}{90 c^3} T^4 
 ~. 
 \label{EOSmatter}
\end{equation}
This is equivalent to the pressure of radiation which has the same
equation of state $w_{\rm matter}=1/3$, since the equation of state for the gas of
 massless ``relativistic'' (quasi)particles
does not depend on whether $c$ is the speed of light or the speed of sound. Since
there is no external pressure, the sum of partial pressures of the condensed-matter 
vacuum and quasiparticles (matter) must be zero, and  one obtains the non-zero vacuum
energy  obeying the equation (\ref{SpecialRelativityCondM}) with $w_{\rm matter}=1/3$.

This demonstrates that the vacuum response to matter is universal. 
Moreover, for the hot Universe, where $w_{\rm matter}=1/3$,  the vacuum
energy density appears to be comparable to the energy density of matter.
But we still did not introduce gravity.
 
 \subsection{Vacuum energy in Einstein Universe}
\label{EinsteinUniverse}

In the presence of gravity, i.e. for $G\neq 0$, we can apply thermodynamic
consideration to the stationary Universes, which allows us to obtain the result for
the vacuum energy density  without solving  Einstein equations. In particular, it can
be done for the static  closed Einstein Universe and the Einstein result is reproduced
\cite{VolovikAnnals}:
\begin{equation}
\epsilon_{\rm vac}= {1\over 2} \epsilon_{\rm matter} (1+3w_{\rm matter})~.
\label{EinsteinSolution}
\end{equation} 
Now, in the presence of gravity, even in the cold Universe, where $w_{\rm matter}=0$,  
the vacuum energy density is comparable to the energy density of matter. This is due
to the response of the vacuum energy to the gravitational field. 
The discussion of the vacuum energy and cosmological constant in the Einstein
Universe is in Apppendix.

\subsection{No response to local perturbation}
\label{Lamb}

There can be other perturbations of the quantum vacuum,  which lead to
the non-zero vacuum energy, such as boundaries which provide the Casimir
effect.  We are almost sure that the Casimir energy emerging between the
neutral  perfectly conducting  plates is gravitating, though
it is still too far from the experimental check.   If so, then one may
ask, why the zero-point energy  in the space between the plates
gravitates and not  the zero-point energy outside the plates. The related
question is why the zero-point energy gravitates in the environment of
the atom  (Lamb shift) and not in vacuum
\cite{Polchinski}. The answer is simple,  the local perturbation of the
vacuum (by an atom or by plates) does not change the pressure at
infinity, and thus the cosmological constant is not perturbed by local
perturbations. 

Actually the answer to these questions has been
given by Einstein.
In Ref. \cite{bib1}, Einstein noted 
that the $\Lambda$-term must be added to
his equations if the density of matter in the Universe is non-zero 
in average.
 In particular, this means that $\Lambda=0$ if matter in the Universe
is so inhomogeneous that its average over big volumes $V$ tends to zero. 
We just discussed the special case when this condition is satisfied, 
i.e. when the  perturbation occupies the finite region of the infinite
Universe.  This is one more example of remarkable Einstein intuition. 

\subsection{Vacuum perturbations in our Universe}
\label{VacuumPerturbations}

In the above simple cases of static perturbations of the vacuum we  found
that the vacuum energy density is on the order of the matter density.
This  gives some hint on how to solve the coincidence problem in
cosmology. 

In our expanding Universe there are several factors which  lead
to non-zero contributions to vacuum energy: 
expansion with the Hubble parameter $H$; gravitating matter with  the
energy density
$\epsilon_{\rm matter}$; temperature $T$; possibly the curvature  $1/R^2$
(experiments demonstrate that our Universe is most probably flat); etc. 
The effect of  the static perturbations  (from curvature,  temperature
and matter) can be easily calculated,  using  the macroscopic
thermodynamic analysis. For the time-dependent perturbations there is no
simple recipes,   because the purely thermodynamic consideration fails to
work in the non-equilibrium situation, 
while the dynamics of the vacuum energy and thus of the 
cosmological constant is not known and most probably is not universal, 
i.e. it may depend on the details of microscopic physics.
We are only able to make the order of magnitude
estimations.

Then one finds that the above perturbations of the vacuum lead  to the
non-zero vacuum energy expressed through the macroscopic parameters $G$,
$T$, $R$, $H$, $\dot H$. 
 The corresponding
contributions to the vacuum energy density  are  proportional toÊ
$H^2 /G$, $\epsilon_{\rm matter}$, $T^4$, ${\dot H}/G$, $1/GR^2$, etc. 
These are the natural macroscopic scales for the cosmological constant,
which are always extremely small compared to the naive estimation of the
vacuum energy
$E_{\rm Planck}^4$ expressed in terms of the
microscopic Planck scale. How the natural value of the  cosmological
constant emerges in the Einstein static Universe is discussed in Appendix.

The dynamics of the cosmological constant is a real problem. 
But there is nothing unnatural with this problem, 
and there are no big puzzles related to cosmological constant: 
The present cosmological constant is small because 
all the perturbations of the vacuum are small, 
and it has the right order of magnitude.

\section{Conclusion}
\label{Conclusion}

Unbearable lightness of space-time, i.e.  that the cosmological term is
too small compared to ``its natural value'', is the typical myth.  A
simple exercises with the vacuum energy of the many-body quantum system
considered here demonstrate that the natural value of the energy density
of the vacuum is determined not by the microscopic (Planck or other)
energy scale cut-off, $E_{\rm Planck}^4/c^3$, but by the macroscopic
parameters of the system. 

There are several ways of how to calculate the  vacuum energy in
condensed matter: Ê(i) using consideration in terms zero-point energy of
the effective bosonic quantum fields, or in terms of the energy of Dirac
vacuum  of the effective fermionic quantum fields (practically any
quantum condensed matter, if its temperature is sufficiently low,  is
described by bosonic and/or fermionic quantum fields);  (ii) by
application of the macroscopic laws of thermodynamics; (iii) by exact
calculations of the energy density using the microscopic (atomic) physics.

The naive consideration (i) gives the big vacuum energy,
$\epsilon_{\rm vac}\sim +E_{\rm Planck}^4/c^3$ or  $\epsilon_{\rm
vac}\sim -E_{\rm Planck}^4/c^3$, depending on the vacuum content.   In
crystals and in superfluid
$^4$He, the role of the Planck energy scale is played  by the Debye
energy; $c$ is the speed of sound; and the sign is
$+$, since the energy is obtained by the summation over  the zero-point
energies of the bosonic fields (phonons). In superfluid
$^3$He, $c$ is the (anisotropic) speed of fermions;  the sign is $-$,
since the contribution comes from the occupied negative energy levels of
the Dirac sea.

The consideration (ii)  gives exactly zero energy density, 
$\epsilon_{\rm vac}=0$, if the condensed matter system satisfies  the
following conditions: it is in complete thermodynamic equilibrium;
stationary;
homogeneous,
does not contain excitations (quasiparticles, which play the role  of
matter); does not interact with environment;
is at $T=0$;
the effects of boundaries are neglected.

This follows from the very simple thermodynamic argument, known  as
Gibbs-Duhem relation. If all these conditions (except for the requirement
of non-interaction with the environment) are satisfied, one obtains the
universal equation of state, 
$\epsilon_{\rm vac}=-p_{\rm vac}$. It is valid both for the vacuum  of
relativistic quantum fields and for the ground state of non-relativistic
quantum condensed matter systems.

If in addition there is no interaction with environment,  the external
pressure is zero and thus the vacuum energy density $\epsilon_{\rm
vac}=0$.Ê

There is a clear contradiction between the results of semi-microscopic 
consideration  (i) and macroscopic consideration (ii). 

The pure microscopic consideration (iii) confirms the zero result  of the
macroscopic approach (ii) under the same conditions as it was obtained
from the macroscopic physics. The microscopic physics also shows how the
contradiction between the naive approach (i) and the macroscopic physics
(ii) is resolved. In addition to the phonon degrees of freedom, the
condensed matter system contains the high-energy degrees of freedom which
cannot be described in terms of the long-wavelength acoustic fields
(acoustic phonon fields are only determined for the wavelength bigger
that the inter-atomic distance). It appears that the high-energy atomic
degrees of freedom automatically adjust themselves in such a way, that
the vacuum as a whole obeys the macroscopic thermodynamic relations. In
the perfect vacuum (satisfying  the above conditions), the contribution
of the high-energy degrees of freedom to the vacuum energy density
automatically compensates the contribution of the sub-Planckian effective
quantum fields to satisfy the zero pressure condition. Thus the
ultraviolet divergence of the contribution of the quantum fields is
absolutely irrelevant for the vacuum energy: the macroscopic effect of
compensation does not depend on the cut-off.

Since the macroscopic thermodynamic consideration does not depend on 
details of the microscopic (trans-Planckian) physics, the same reasoning
should is applicable to the vacuum of relativistic quantum fields.  The
only assumption made is the mere existence of the microscopic
(trans-Planckian) degrees of freedom. If they exist, they must obey the
macroscopic laws and thus they will be automatically arranged to satisfy
the zero pressure condition in the absence of the interaction with the
environment.

If the vacuum is perturbed, the microscopic degrees of freedom are 
automatically rearranged to satisfy the new equilibrium. If the
perturbation is caused by the phase transition from one vacuum to another
(say, from false to true vacuum), the microscopic degrees of freedom will
be slightly deformed to adjust to the new vacuum state. As a result,
after the equilibrium is reached, the zero value of the vacuum energy
density is restored in the new vacuum.

One can also find that the readjustment of the microscopic parameters 
after the transition is extremely small. For example, in superfluid
$^3$He, the relative change of the inter-atomic distance $a$ (Planck
length) after the superfluid transition is of order
$\delta a/a\sim T_c^2/E_F^2\sim 10^{-6}$, where $E_F$ is  the microscopic
energy  (Fermi energy) and
$T_c$ is the transition temperature. In a similar way, after  the
electroweak transition,  the relative change of the Planck scale and thus
the relative change of the Newton constant
$G\propto 1/E_{\rm Planck}^2$ are of order  $\delta G/G\sim M_{W}^4/
E_{\rm Planck}^4$, whereÊ$M_{W}$ is the gauge boson mass.Ê
 
If the perturbations are caused by matter, boundaries, curvature, 
temperature, expansion, etc., the non-zero vacuum energy emerges. One can
easily calculate the response of the vacuum energy density to the static
perturbations. Then one finds  that vacuum energy density is proportional
to perturbations of the vacuum, and is determined by macroscopic
parameters, rather than microscopic. The same occurs for the
time-dependent perturbations. Though in this case we are not able to
calculate the vacuum response exactly, since we do not know the dynamics
of the vacuum energy, we can expect that the order of magnitude
estimation in terms of the macroscopic parameters remains correct.

In our expanding Universe there are different perturbations of  the
quantum vacuum: expansion with the Hubble parameter $H$; gravitating
matter with the energy density
$\epsilon_{\rm matter}$; temperature $T$; possibly a small curvature 
$1/R^2$; etc. One can estimate that these perturbations lead to the
non-zero vacuum energy expressed through the macroscopic parameters $G$,
$T$, $R$, $H$, $\dot H$. The corresponding contributions to the vacuum
energy density  are  proportional toÊ
$H^2 /G$, $\epsilon_{\rm matter}$, $T^4$, ${\dot H}/G$, $1/GR^2$, etc. 
These are the natural macroscopic scales for the cosmological constant,
which are always extremely small compared to the naive estimation of the
vacuum energy
$E_{\rm Planck}^4$ expressed in terms of the
microscopic Planck scale (see also Appendix).

Concerning  the absolute value of vacuum energy, some people think  that
it only has significance for gravity, otherwise it can be shifted by the
redefinition of the zero energy. From the condensed matter examples one
can find that the properly defined energy density, which is relevant for
the vacuum of quantum fields, does not depend on the choice of the zero
level. The overall shift of the energy does not change
$\epsilon_{\rm vac}$: the vacuum energy density (as well as pressure)  is
the intensive thermodynamic quantity (not extensive). It is the real
thermodynamic quantity  well defined in any condensed matter independent
of whether the effective gravity emerges or not in the systems.

The vacuum response to the time dependent perturbations and the  dynamics
of the vacuum energy and cosmological constant is the real problems. Even
in the many-body systems, where we know all the microscopic physics, the
non-equilibrium dynamics is the complicated subject. But there is nothing
unnatural with the problem of the dynamical adjustment of the vacuum
energy to perturbations of the vacuum. Let us look at the general
features of the adjustment mechanism discussed in Ref. \cite{Dolgov}:

\noindent 1. Some compensating agent must exist.

\noindent In our approach the compensating agent 
is the reservoir of the trans-Planckian degrees of freedom of  quantum
vacuum, with thermodynamic compensation occurring without fine tuning.

\noindent 2. Quite natural to expect that the compensation is not 
complete and the resulting vacuum energy density  is close to the
critical energy density.

\noindent This is also naturally fulfilled in our approach,  where the
compensation is not complete if the vacuum is not perfect.

\noindent 3. A realistic model is needed.

\noindent In our approach this must be the phenomenological   theory
which describes the dynamics of the trans-Planckian degrees of freedom.
In condensed matter it is the dynamics of the density of the underlying
atoms, $n\propto E_{\rm Planck}^3$, which is responsible for the
relaxation of the vacuum energy density to its equilibrium value. The
dynamics of the 3D or 4D density $n$ of the underlying `atoms' of the
vacuum, which is proportional to $E_{\rm Planck}^3$ or $E_{\rm Planck}^4$
correspondingly, can be probably constructed as the dynamic extension of
the equations discussed in Appendix. This will allow us to consider the
decay of our Universe to the perfect equilibrium vacuum state with
Minkowski space-time, zero $\Lambda$ and no matter.

In conclusion, there are no unbearable contradictions between  the
theoretical estimate for the vacuum energy and the observed dark energy.
In spite of the ultraviolet divergence of the zero-point energy,  the
natural value of the vacuum energy is comparable with the observed dark
energy. That is why the vacuum energy is the plausible candidate for the
dark energy.

\section*{Acknowledgments}

I thank James Bjorken, Sean Carroll, Alexander Dolgov and Frank Wilczek 
for e-mail correspondence. This work is supported in part by the Russian
Ministry of Education and Science through the Leading Scientific School
grant $\#$1157.2006.2,  and by the European Science Foundation  COSLAB
Program.

\section*{Appendix. Cosmological constant in Einstein Universe}
 
Let us illustrate how the natural macroscopic value of  the
vacuum energy and cosmological constant emerges in the Einstein
static closed Universe. We start with the action:
\begin{equation}
S=S_{\rm E}+S_\Lambda+S_{\rm matter}~~.
\label{EinsteinAction1}
\end{equation}
Here $S_{\rm matter}$ is the matter action; $S_{\rm E}$  is the
Einstein curvature action; and $S_\Lambda$ 
is the action which gives rise to the cosmological term.  We
shall treat the vacuum energy in the same way as in condensed
matter, i.e. by describing the  microscopic (Planckian) degrees
of freedom of the quantum vacuum in terms of the ground state of
the system of some constituent particles with particle number
density $n$ and the  chemical potential $\mu$ -- the Lagrange
multiplier responsible for the conservation of the particle
number (though the physical meaning of quantities $n$ and $\mu$ may be
different, see below). This approach assumes that the microscopic physics
contains the principal collective mode, which bears the information on
the vacuum energy. The mass of this mode has the Planck energy scale.

Following the condensed matter experience, we choose the following
phenomenological action for the real scalar field $n$:
\begin{equation}
 S_\Lambda=
-\int d^4x\sqrt{-g}\left(\epsilon(n) -\mu n + \gamma(n)
g^{\mu \nu}\partial_\mu n\partial_\nu n\right)~.
\label{EinsteinAction3}
\end{equation}

In the Einstein curvature action the Newton constant $G$
depends on $n$:
\begin{equation}
S_{\rm E}= -{1\over
16\pi }\int d^4x \frac{1}{G(n)}\sqrt{-g}{\cal R}~.
\label{EinsteinAction2}
\end{equation}
The total action is similar to the
scalar-tensor theory of gravity with spin-0 scalar field $n$
(see e.g. \cite{Uzan}). The only difference is that the
potential for the $n$-field contains the Lagrange multiplier
$\mu$.

 The introduced quantities have the
natural Planck scales, for example: 
$\mu   \sim E_{\rm Planck}$; $ \epsilon \sim E_{\rm Planck}^4$;
$n\sim E_{\rm Planck}^3$;
$\gamma \sim  E_{\rm Planck}^{-4}$; $G^{-1}\sim  E_{\rm
Planck}^{2}$; and $M\sim  E_{\rm Planck}$ where $M$ is the mass
of the $n$ field. 
In this presentation, the quantity $n\sim E_{\rm Planck}^3$
plays the role of the number density of the quantum modes in the
vacuum. In condensed matter, the number of modes per volume is
finite (it is 3 $\times$ number density of atoms): the momentum
and energy of each atom is not constrained in the excited state,
but the energy, momentum and number of the modes which form the
ground state of, say, crystal, are limited. Then  $\mu$ serves
as the Lagrange multiplier responsible for the conservation of
the number of modes in the quantum vacuum.

In principle, the physical
meaning of
$\mu$ and $n$ and their dimensions can be different. For
example, one can choose the following sets, which are  more compatible
with Lorentz invariance: (1) $ \mu \sim \epsilon
\sim E_{\rm Planck}^4$,
$n\sim  E_{\rm Planck}^{0}$,
$\gamma \sim  E_{\rm Planck}^{2}$, $M\sim  E_{\rm Planck}$; and (2)
$ n \sim \epsilon
\sim E_{\rm Planck}^4$,
$\mu \sim  E_{\rm Planck}^{0}$,
$\gamma \sim  E_{\rm Planck}^{-6}$, $M\sim  E_{\rm Planck}$.  In the
first set, $\mu$ has dimension of vacuum energy and cosmological
constant. In the latter set it is the 4D + 4D phase space which is
conserved, with dimensionless Lagrange multiplier $\mu$. 
Also for
generality, the equation must be added which allows $\mu$ to relax. But
here it is not important since we are  looking for the stationary
solutions corresponding to Einstein Universes. 

 The variation over the metric $g^{\mu\nu}$ gives the
Einstein equations:
\begin{equation}
-{1\over 8\pi G}\left( R_{\mu\nu}-{1\over
2}{\cal R}g_{\mu\nu}\right)+ (\epsilon(n) -\mu n) g_{\mu\nu}
+T^{\rm M}_{\mu\nu} =0~,
\label{EinsteinEquation1}
\end{equation}
where $T^{\rm M}_{\mu\nu}$ is the
energy-momentum tensor for matter;  and $\epsilon(n) -\mu n$
plays the role  of the cosmological constant $\Lambda$. We do
not take into account the gradients of
$n$ because of the big Planck-size mass  of the $n$-field, 
while we are looking for the homogeneous solution of
equations.
 The variation over   $n$ gives
\begin{equation}
\frac{d\epsilon}{dn}-\mu+\frac{1}{16\pi}\frac{dG^{-1}}{dn}{\cal
R}=0~.
\label{nEquation1}
\end{equation}
For the Einstein Universe the solution of the Einstein
equations gives
\begin{eqnarray}
\Lambda\equiv \epsilon(n) -\mu n =\frac{1}{2}
\rho_{\rm matter}(1+3w_{\rm matter})~,
\label{LambdaGeneralEinstein}
\\
 \frac{1}{4\pi GR^2}=\rho_{\rm matter}(1+w_{\rm matter})~,
\label{Radius}
\end{eqnarray}
where $R$ is the radius of the Universe,  and $P_{\rm
matter}=w_{\rm matter}\rho_{\rm matter}$ is the equation of
state for matter. Equation (\ref{nEquation1}) with ${\cal
R}=-6/R^2$, and
$dG^{-1}/dn=\beta G^{-1}/n$ with $\beta$ of order unity gives
\begin{equation}
\frac{d\epsilon}{dn}-\mu=\frac{3\beta}{8\pi GnR^2}~.
\label{nEquation2}
\end{equation}
These 3 equations (\ref{LambdaGeneralEinstein}--\ref{nEquation2}) 
express $n$, $\mu$ and $R$ in terms of matter density.

Let $n_0$ and $\mu_0$ be the values at $R=\infty$ when
$\rho_{\rm matter}=0$, which corresponds to the perfect  vacuum
state. These quantities obey equations
\begin{equation}
(d\epsilon/dn)\vert_{n_0}=\mu_0 ~,~\epsilon(n_0) =\mu_0 n_0
~.
\label{Vacuum}
\end{equation}
These are exactly the equations for the ground state of 
condensed matter system in the absence of environment, i.e. at
zero external pressure (see Sec. 3.3 in Ref.\cite{VolovikBook}). The
cosmological constant and vacuum energy in this pure and non-disturbed
vacuum are zero as expected from the thermodynamic arguments:
\begin{equation}
 \Lambda(n_0)=\epsilon(n_0) - \mu_0 n_0=0 ~.
\label{ZeroLambdaVacuum}
\end{equation}
 
Gravity and matter perturb the vacuum energy and  $\Lambda$. 
These perturbations are obtained by  expanding
$\epsilon(n)-\mu n$ in the vicinity of the  values $n_0$ and $\mu_0$
in the perfect vacuum:
\begin{equation}
\epsilon(n)-\mu n  \approx 
\frac{1}{2}\epsilon'' ( \delta n)^2-n_0 \delta\mu -\delta n \delta\mu~.
\label{Expansion}
\end{equation}
Then the first-order corrections are:
\begin{eqnarray}
\Lambda\equiv -n_0\delta \mu  =\frac{1}{2}
\rho_{\rm matter}(1+3w_{\rm matter})~,
\label{FirstOrder1}
\\
 n_0\epsilon''\delta n-n_0\delta \mu =\rho_{\rm matter}
(1+w_{\rm matter})~,
\label{FirstOrder2}
\end{eqnarray}
or
\begin{eqnarray}
 \frac{\Lambda}{\Lambda_{\rm Planck}}\sim \frac{\delta
\mu}{\mu_0}  =-\frac{ 
\rho_{\rm matter}(1+3w_{\rm matter})}{2n_0^2\epsilon''}~,
\label{FirstOrder3}
\\
  \frac{\delta G}{G}=- \beta \frac{\delta n}{n_0}=
\frac{ 
\rho_{\rm matter}(1-w_{\rm matter})}{2n_0^2\epsilon''}\sim  \frac{\Lambda}{\Lambda_{\rm Planck}}~.
\label{FirstOrder4}
\end{eqnarray}
Here $\Lambda_{\rm Planck}\sim E_{\rm Planck}^4$ (for any choice
of the field $n$, one has $n_0^2\epsilon''\sim E_{\rm
Planck}^4$); it is the
``natural'' value of the cosmological constant, as it follows
from the naive estimation of the vacuum energy. The  true cosmological constant $\Lambda$ in Eq.(\ref{FirstOrder1}), which 
naturally emerges from the macroscopic physics, is much smaller.  It does not depend on the details of the microscopic (trans-Planckian) physics.

On the other hand, the value of the Newton constant $G$ is
almost completely determined by the microscopic physics of the vacuum: the
corrections to $G$ caused by matter and gravity in Eq.(\ref{FirstOrder4})
are extremely small. This rules out any time dependence of $G$ during the
observational cosmological time.

Equation (\ref{FirstOrder3}) demonstrates that, though the microscopic (Planck)  degrees
of freedom are described by the natural Planck scales, the
macroscopic physics is governed by the macroscopic scales.
In particular, in the Einstein Universe the natural value of
$\Lambda$ is determined by the energy density of matter 
$\rho_{\rm matter}$ and the radius $R$ of the Universe, rather
than by the Planck energy scale.

What should be done next? First the dynamical
equations must be extended to describe
the relaxation of the parameter $\mu$ (and thus of the cosmological
constant
$\Lambda$), including the readjustment of these parameters to new
equibrium after the phase transition.  Then the principal  mechanism of
relaxation must be found which drives the non-equilibrium Universe to the
final equilibrium state with zero
$\Lambda$ and without matter (maybe after several cicles). In condensed
matter, one of the possible scenarios of the decay of the
non-equilibrium vacuum is the Suhl instability \cite{Suhl,LvovReview}. It
is the parametric instability  of the precessing magnetization (analog of
the time-dependent vacuum state) with respect to excitations of pairs of
spin waves (analog of radiation of pairs of photons or/and gravitons
in the process of reheating).


\end{document}